\documentclass[12pt]{article}
\usepackage[cp1251]{inputenc}
\usepackage[english]{babel}

\usepackage{graphicx}
\usepackage{amsfonts}
\usepackage{amssymb}
\usepackage{amsmath}
\usepackage[colorlinks,urlcolor=blue]{hyperref}

\usepackage{mathrsfs}
\usepackage{listings}

\begin{document}

\title{\textbf{Orbitron. Part I.\\
Stable orbital motion of \\
magnetic dipole \\
in the field of permanent magnets}}

\author{Stanislav S. Zub$^{*}$}
\date{}
\maketitle

\tableofcontents

\makeatletter

\let\oldthefootnote\thefootnote
\renewcommand{\thefootnote}{\fnsymbol{footnote}}
\footnotetext[1]{Faculty of Cybernetics. Taras
Shevchenko National University of Kyiv. \\
Glushkov boul., 2, corps 6. UA-03680, Ukraine. E-mail: \url{stah@univ.kiev.ua}}
\let\thefootnote\oldthefootnote

\makeatother

%
\inputencoding{cp1251}
\newcommand{\s}[1]{\ensuremath{\boldsymbol{\sigma}_{#1}}}
\newcommand{\bsym}[1]{\ensuremath{\boldsymbol{#1}}}
\newcommand{\w}{\ensuremath{\boldsymbol{\wedge}}}
\newcommand{\lc}{\ensuremath{\boldsymbol{\rfloor}}}
\newcommand{\rc}{\ensuremath{\boldsymbol{\lfloor}}}

\thispagestyle{empty}

\newpage
\section{Introduction}
\label{introduction}
\bigskip
The problem of magnetic configurations stability has a long history.
In 1600 W.Gilbert published a treatise ‘‘On the Magnet, Magnetic Bodies,
and the Great Magnet of the Earth’’ where he proposed
that magnets can form the noncontact stable systems.

Since then a number of world known scientists, for example, Newton,
Earnshaw, Heisenberg, Kapitsa, Braunbeck, Tamm,
Ginzburg made their contributions to the study of this problem.

The problem of magnetic equilibrium stability can be naturally divided
into two tasks of static and dynamic equilibrium.

Unfortunately, in both of these scientific fields a number of prejudices and errors appeared
which are not entirely solved until now.

Concerning the static equilibrium, this resulted in unjustified transference of the conclusion
from Earnshaw theorem about systems instability in electrostatics into the field of magnetic phenomena.
Partly this error was eliminated in the light of magnetic levitation experiments conducted by Braunbeck
and Kapitsa-Arkad'ev (i.e. in combination both of magnetic and gravity forces).

Studies [1] as well as dissertation [2] were devoted to solving the problem
of static equilibrium of bodies, which interact only via magnetic force.

Particular prejudices have also penetrated the problem of dynamic stability
equilibrium in magnetic systems.

At the dawn of the nuclear age, in the years when research of an atomic nucleus has been thriving,
magnetic interactions were considered as a possible mechanism of keeping particles in the nucleus.
In 1941- 1947 Tamm and Ginzburg have shown that in the case of two interacting magnetic dipoles
the orbital motion is impossible, due to the particles falling down the center both in classical
and in quantum mechanics [3]. In physics this fact was called ‘‘problem $1/r^3$’’ and together
with Earnshaw theorem, extended to the case of magnetostatics,
resulted in opinion of ‘‘global instability of the magnetic systems’’ for a long time.

However even after these results famous physicists,
J.Schwinger [4] for example, had their interest in the magnetic model of matter.

New splash of interest to the problem of dynamic equilibrium
in the magnetic systems resulted in creation by Roy Harrigan  a levitron in 1983.
Internet provided wide possibilities for popularization of this unusual toy,
and also other experiments with magnetic bodies [5].

Against this background the importance of theoretical
and experimental research conducted by V. Kozorez
almost ten years before, in 1974 [6,7]
seemed to attract unfairly small interest among the world physics community.

V. Kozorez tried to develope the known idea of the solution
of the magnetic systems stability problem based on the consideration
that magnetic particles are extended presented by Heisenberg
in the twenties of last century.

He succeeded in building the experimental prototype where a small magnet
accomplished quasiorbital motion up to 6 minutes in duration (this prototype,
in contrast to levitron was not patented).

As we have already noted in [8,9] his theoretical research had rather estimating character,
because adequate mathematical apparatus [10-13] for studying the stability of such systems
has not been developed yet.

In particular, the condition of stability that he has got for the system analogous
to the system considered in this article is only one of the three sufficient conditions for stability.
This condition gives the exact expression which reflects the well known fact,
that on a considerable distance any magnetic system presents a dipole.

In respect of the experiment per se,
from the philosophical point of view an experiment as such
or computational modeling in principle cannot prove the dynamic stability,
but can only give certain reasons in support of the stability [14].

For the first time the strictly analytical proof of
an orbital motion stability in magnetic systems is given in [8,9].
Analytical conditions for the stability of the system formed by
two ‘‘magnetic dumbbells’’ have a complicated form and values of parameters
for the system’s stability area were determined numerically.

Therefore it makes sense to consider a simpler magnetic system
that we call Orbitron for convenience.
Here we analytically prove not only the existence of stable orbits
but also stability conditions which have simple physical meaning for this system.

In this system a movable body is a small permanent magnet with an axial symmetry.
Its interaction with magnetic field is described by magnetic dipole approximation,
and its motion obeys the laws of rigid body motion.

This differs from the model accepted in work [15].
We do not use the analogy originated from the attempts of classical description
of such quantum-mechanical parameter as particle spin.

Thus one of tasks of this article is to provide the motion equations of such
magnetic rigid body for the systems like Kozorez’s prototype and levitron.

Here, for the mathematical model we call Orbitron,
we give accurate analytical prove of the existence of stable orbital motion
of a small magnetized rigid body, described as a magnetic dipole based on the theorem from [12].
The sufficient conditions of stability in this model have a simple form
allowing clear physical interpretation.

\newpage
\section{Hamiltonian formalism for magnetic dipole in the axisymmetrical magnetic field }
\label{Hamilton}
\bigskip

Let’s consider the Hamiltonian dynamics of ‘‘small’’ rigid body
in the axisymmetrical magnetic field assuming magnetic dipole approximation.
Such field can be created by cylindrical magnets, solenoids,
current-carrying rings and other objects with axial symmetry along $z$ axis.

The variant of such formalism can be obtained from the formalism of work [16]
by the limiting process $m_1\longrightarrow~\infty$.
Thus we consider the first body immobile and $z$ axis oriented.
Then its dynamical variables disappear from consideration or become model parameters.
Therefore the group of symmetry of the task converges to $SO(1)$.

For constructing Hamiltonian dynamics based on Poisson structures
it is necessary to specify a Poisson manifold and also kinetic
and potential energy of the system.

Poisson manifold of Orbitron is the direct product of Euclidean spaces

\[ P = R^3_x\times R^3_p\times R^3_\nu\times R^3_n
\leqno(1)\]
with Poisson brackets for correspondent generatrix.

Generatrix for Orbitron will be: $x_i$ - dipole coordinates;
$p_i$ - its components of momentum (orbital motion);
$n_i$ - components of  dipole intrinsic moment of momentum;
$\nu_i$ - components of directing unit vector of dipole’s axis of symmetry.

Nonzero Poisson brackets between generatrix on $P$ look like
\[ \begin{cases}
  \{x_i,p_j\} = \delta_{ij}; \\
  \{n_i,\nu_j\} =\varepsilon_{ijk}\nu_k; \quad \{n_i,n_j\} = \varepsilon_{ijk} n_k.
\end{cases}
\leqno(2)\]

Casimir functions of this Poisson structure that are easily checked
will be $\vec{\nu}{\ }^2=1$ and $(\vec{\nu},\vec{n})=const$.

System Hamiltonian we write down in the form:
\[
h=T + U(r,c^{'},c^{''},c^{'''}),
\leqno(3)\]
where
\[ U = -(\vec{\mu}\cdot\vec{B})
\leqno(4)\]
\[ \vec{B}(\vec{r}) = B_r(r,c')\vec{e}_r + B_z(r,c')\vec{e}_z,
\leqno(5)\]
\[
  \begin{cases}
     r = |\vec{r}|;\\
     \vec{e}_r = \vec{r}/|\vec{r}|;\\
     c' = \vec{e}_z\cdot\vec{e}_r = x_3/r;\\
     c'' = \vec{\nu}\cdot\vec{e}_r;\\
     c''' = \vec{e}_z\cdot\vec{\nu} = \nu_3.
  \end{cases}
\leqno(6)\]

As usual, kinetic energy of movable body (dipole) consists of
kinetic energy of both translational and rotational motions [16,8].
\[ T(p^2,\vec{n}^2)= \frac1{2 M} p^2 +\frac{\alpha}2\vec{n}^2,
\]
where $M$ -- dipole mass;
    $\alpha=\frac{1}{I_\bot}$
(as well as before we suppose, that $I_1=I_2=I_\bot$, where $I_1,I_2,I_3$ --
intrinsic moments of the body’s inertia).

Get the system of motion equations for magnetic dipole in axisymmetrical magnetic field:
\[ \begin{cases}
  \dot{\vec{r}} = \vec{p}/M; \\
  \dot{\vec{p}} = -{\partial}_{r}U\vec{e}_r -
  \frac{1}{r}({\partial}_{c^{'}}U P_{\bot}^e(\vec{e}_z)
  +{\partial}_{c^{''}}U P_{\bot}^e(\vec{\nu}));\\
  \dot{\vec{\nu}} =  \alpha(\vec{n}\times\vec{\nu});\\
  \dot{\vec{n}} = -\vec{\nu}\times
                   (\vec{e}_r\partial_{c^{''}}
                  + \vec{e}_z\partial_{c^{'''}})U,
\end{cases}
\leqno(7) \]
where $P_{\bot}^e$ -- projector on the plane perpendicular to the vector $\vec{e}_r$,
i.e.
$P_{\bot}^e(\vec{e}_z)=\vec{e}_z-c^{'}\vec{e}_r$ and
$P_{\bot}^e(\vec{\nu})=\vec{\nu}-c^{''}\vec{e}_r$.

Expressions for the force and the force momentum acting on a dipole
in an external magnetic field are well known.
We can show that the second and fourth equations of the system (7)
can be presented in classical representation.

Concerning the second equation in the system (7),
it has been obtained from the standard expression of Hamiltonian formalism
\[ \dot{p_i} = \{p_i, H \} = \{p_i, U \}
   = \partial_r U \{p_i,r\}
   + \partial_{c^{'}} U \{p_i,c^{'})\}
   + \partial_{c^{''}} U \{p_i,c^{''})\}
\leqno(8)\]

For potential energy in form (4) we obtain a classic expression of the force
\[ \dot{\vec{p}} = \{\vec{p}, H \} = \{\vec{p}, U \}
  = - \nabla U =  \nabla (\vec{\mu}\cdot\vec{B})
\leqno(8a)\]

Regarding the fourth equation in the system (7),
the potential energy of a dipole is described by formula (4)
in the axisymmetrical magnetic field which is described by formula (5),
therefore we obtain
\[ (\vec{e}_r\partial_{c^{''}}
  + \vec{e}_z\partial_{c^{'''}})U
  = -\mu \vec{B}
\leqno(9)\]

Then we get the last equation in the system (7) in usual classical representation
\[ \dot{\vec{n}} = \mu\vec{\nu}\times\vec{B}
                 = \vec{\mu}\times\vec{B}
\leqno(10)\]

Now the system of equations (7) can be written in the form:
\[ \begin{cases}
  \dot{\vec{r}} = \vec{p}/M; \\
  \dot{\vec{p}} = \nabla (\vec{\mu}\cdot\vec{B});\\
  \dot{\vec{\mu}} = (\vec{n}\times\vec{\mu})/I_\bot;\\
  \dot{\vec{n}} = \vec{\mu}\times\vec{B},
\end{cases}
\leqno(7a) \]

A few remarks are necessary regarding the systems of motion equations (7,7a).

1. Both systems are correct in the quasi-stationary electromagnetic field approximation [17,18].
This approximation is characterized by the possibility to neglect
the finiteness of electromagnetic disturbances propagation speed
and displacement current in the range of the system
and calculate magnetic fields using formulas of magnetostatics.

2. The system of equations (7) uses the concept of magnetic potential energy,
which is incident to long-range action conception in classic mechanics.
As just was mentioned, this is possible in quasi-stationary approximation.
The chosen form of the potential energy, as in formula (3),
describes not only dipoles but also wide enough class of
axisymmetrical magnetic bodies.

3. The system (7a) corresponds to the concept of short-range interactions
in the electromagnetic field theory.
Therefore these equations are obviously valid not only for
the axisymmetrical magnetic field but also describe the motion of a dipole
in {\it an arbitrary} external magnetic field.

4. We consider a magnetic dipole, as a small magnetized rigid body
with axial symmetry as in levitron for example.
Equation (1) in work [15] in this case could not replace
the third and fourth equations of the system (7a).
This distinguishes our mathematical model
from that accepted in work [15].

\newpage
\section{Mathematical model of Orbitron }
\label{Model}
\bigskip

Not all axisymmetrical magnetic fields can create the possibility
for a stable orbital motion of a magnetic dipole.
For example, the field of magnetic-dipole type results in
‘‘problem $1/r^3$’’ mentioned in introduction.
Therefore, it may be useful to use Heisenberg’s hypothesis
about the possibility of stable magnetic configurations
with magnetic extended bodies (see also [7]).

Here we offer the following model of Orbitron.

Put two magnetic unlike poles on axis $z$ at points $\pm h$.
These poles create the axisymmetrical magnetic field
in which a magnetic dipole is moving.
We assume that stable orbital motion of the system
is possible under certain parameters.

It is important to give some explanation here.
Equations of magnetostatics do not suppose
the existence of isolated magnetic charges.
However, the field outside a thin solenoid, for example
(the same for the thin cylindrical magnet)
will coincide with high accuracy with the field of two poles [19].
On the other hand, the field {\it inside the solenoid} not only does
not coincide with charges field but also opposite in sign,
so that the flow through the unbounded surface embracing
only one pole is equal to zero, as required by magnetostatics equations.

It is assumed that the dipole moves a sufficient distance from the poles of the magnet,
which is the source of the field, and the model of two magnetic charges describes
the field with high accuracy.

Thus, magnetic field in the system has the form of sum of the coulomb fields of two charges $\pm\kappa$:
\[ \vec{B}(\vec{r})
  = \sum_{\varepsilon=\pm 1} \vec{B}_{\varepsilon}(\vec{r}), \qquad
   \vec{B}_{\varepsilon}
  = \frac{\mu_0}{4\pi}\varepsilon\kappa
    \frac{\vec{r}-\varepsilon h \vec{e}_z}{|\vec{r}-\varepsilon h \vec{e}_z|^3}.
\leqno(10)\]
where each of the fields $\vec{B}_{\varepsilon}$,
and consequently the total field can be presented by formula (5).

Then for the potential energy of a dipole in the magnetic field we get the expression
\[ U(r,c',c'',c''')
   = -\frac{\lambda_0}{4\pi}\sum_{\varepsilon=\pm 1}\varepsilon U_\varepsilon(r,c^{\ '},c^{\ ''},c^{\ '''}),
   \quad \lambda_0 = \mu_0\kappa\mu
\leqno(11)\]
where
\[ U_\varepsilon(r,c^{\ '},c^{\ ''},c^{\ '''})
   = \frac{r c^{\ ''} - \varepsilon h c^{\ '''}}{R_\varepsilon(r,c^{\ '})^3}
\leqno(12)\]
and
\[ R_\varepsilon(r,c^{\ '}) = \left(r^2 - 2\varepsilon h r c^{\ '} + h^2\right)^{1/2}
\leqno(13)\]

Let’s show the first derivatives of the function $U_\varepsilon$:
\[\partial_r U_\varepsilon
= \frac{c^{''}}{R_\varepsilon(r,c^{\ '})^3}
- \frac{3 \left(r c^{''} - \varepsilon h c^{'''} \right)
   \left(r - \varepsilon h c^{'} \right)}{R_\varepsilon(r,c^{\ '})^5}
\leqno(14) \]

\[\partial_{c^{'}} U_\varepsilon
= \frac{3\left( r c^{''}
- \varepsilon h c^{'''} \right) \varepsilon h r}{R_\varepsilon(r,c^{\ '})^5}
\leqno(15) \]

\[\partial_{c^{''}} U_\varepsilon
= \frac{r}{R_\varepsilon(r,c^{\ '})^3}
\leqno(16) \]

\[\partial_{c^{'''}} U_\varepsilon
= -\frac{\varepsilon h}{R_\varepsilon(r,c^{\ '})^3}
\leqno(17) \]

\newpage
\section{Example of stable orbital motion }
\label{Stable}
\bigskip

The main aim of this work (i.e. Part I) is to prove the existence of stable orbital motion
in the systems of bodies, which interact only by magnetic forces.
The example of such system is described in a section 3,
and example of a stable orbit will be the circular orbit in plane $z=0$.

\subsection{Relative equilibrium }
\label{RE}

A special role in orbital motions stability of Hamiltonian systems plays
the so-called {\it relative equilibrium} [11,12],
i.e. such trajectories of the dynamic system which simultaneously
are one-parameter sub-groups of the system’s invariance group.

As it has been already mentioned, the invariance group of Orbitron is $SO(1)$.
Every one-parameter sub-group of this group is characterized
by the intrinsic rotational angular velocity $\vec{\omega}=\omega \vec{e}_z$.
For our problem the rate of change of any physical value $\vec{v}$
along the orbit of the sub-group will be set by the formula $\dot{\vec{v}}=\vec{\omega}\times\vec{v}$.

Therefore, for the relative equilibrium to exist the following relationships must hold
\[
\begin{cases}
  \dot{\vec{r}} = \omega (\vec{e}_z\times \vec{r}); \\
  \dot{\vec{p}} = \omega (\vec{e}_z\times \vec{p}); \\
  \dot{\vec{\nu}} =  \omega (\vec{e}_z\times \vec{\nu}); \\
  \dot{\vec{n}} = \omega (\vec{e}_z\times \vec{n}).
\end{cases}
\leqno(18) \]

We show that a dynamic orbit for which these relationships are satisfied exists.
Examine an orbit, spatially located in the $z=0$ plane.
Also suppose that
$\vec{\nu} \parallel \vec{e_z}$ and $\vec{n} \parallel \vec{e_z}$.
Then $c^{'}=c^{''}=0$, $c^{'''}=\pm 1$ along the whole trajectory
and ${\partial}_{c^{'}}U = {\partial}_{c^{''}}U = {\partial}_{c^{'''}}U =0$
as follows from formulas (11,15-17).

So, the third and the fourth equations of the system (7) then hold identically,
the first and the second are reduced to the second order equation:
\[M \ddot{\vec{r}} + \left(\frac{\partial_r U}{r}\right)_{|r=r_0}\vec{r} = 0
\leqno(19) \]
On condition that $(\partial_r U)_{|r=r_0}>0$ equation (19) has solution
corresponding to the motion on the circumference
with radius $r_0$ and frequency, which is determined by relationship
\[\left(\frac{\partial_r U}{r}\right)_{|r=r_0} = \omega^2 M
\leqno(20) \]

Thus, one can prove that the reduced orbit indeed is a relative equilibrium.

Theorem 4.8. in [12] is a suitable instrument for investigating stability
of relative equilibria on Poisson manifolds.
Important advantage of group theoretical methods is that
the functional space of investigation of trajectories
is substituted by investigation of finite-dimensional
vector space of dynamic variables variations in {\it a fixed}
point on the trajectory.
Thus the investigation approach for stability is very similar
to the study of a function’s conditional extremum by Lagrange multiplier method.

\subsection{Choice of supporting point }
\label{Point}

Lets set the point on an orbit of relative equilibrium
\[ z_e =
   \begin{cases}
      \vec{x}_0 = r_0\vec{e}_1; \\
      \vec{p}_0 = p_0\vec{e}_2;\\
      \vec{\nu} = -\vec{e}_3;\\
      \vec{n} = n_0\vec{e}_3;
  \end{cases}
\leqno(21) \]

Notice that we do not fix the sign of the mechanical moment $n_0$,
it can be {\it arbitrary}. As for a sign of $p_0$,
for a positive angular velocity its value will be positive.

Lets show that in supporting point the following relationships hold
\[ \partial_{c'}U_{|z_e} = 0; \qquad \partial_{c''} U_{|z_e} = 0;
\leqno(22)\]

Since in supporting point
\[ \begin{cases}
      c' = 0; \\
      c'' = 0; \\
      c''' = -1;
\end{cases}
\leqno(23) \]
therefore
\[R_\varepsilon(r,c')_{|z_e} = (r^2 + h^2)^{1/2}
\leqno(24) \]

From the expressions of potential energy derivatives (15-16)
\[(\partial_{c'} U_\varepsilon)_{|z_e}
  = \frac{3 h^2  r}{(r^2 + h^2)^{5/2} },\qquad
  (\partial_{c''} U_\varepsilon)_{|z_e}
  = \frac{r}{(r^2 + h^2)^{3/2} }
\leqno(25) \]
notice that both expressions do not depend on $\varepsilon$,
meaning that in sum on $\varepsilon$
(with $\varepsilon$- multiplier) they will give 0.

\newpage
\subsection{Necessary condition of stability and Lagrangian coefficients}
\label{Stability1}
\bigskip

As motion integrals we will take
\[ \begin{cases}
      j_3 = x_1 p_2 - x_2 p_1 + n_3;\\
      C_1 = \frac{\lambda_1}2\vec{\nu}^{2};\\
      C_2 = \lambda_2(\vec{\nu},\vec{n});
   \end{cases}
\leqno(26) \]
where 1st line represents a third conserved quantity of a body
total angular momentum, and the other two are Casimir functions of the system.

Write out the correspondent differentials in $z_e$ point
\[ \begin{cases}
     (\bsym{d}j_3)_{|z_e} = p_0 \bsym{d}x_1 + r_0\bsym{d}p_2 + \bsym{d}n_3;\\
     (\bsym{d}C_1)_{|z_e} = -\lambda_1\bsym{d}\nu_3;\\
     (\bsym{d}C_2)_{|z_e} = \lambda_2(n_0\bsym{d}\nu_3 - \bsym{d}n_3);
\end{cases}
\leqno(27) \]

Efficiency function (adjoined Hamiltonian) looks like
\[ \tilde{H} = T + U - \omega j_3 + \lambda_1 C_1 + \lambda_2 C_2
\leqno(28)\]

The necessary condition of stability in theorem 4.8. [12]
requires the differential of efficiency function to be equal
to zero in a supporting point, i.e. $\bsym{d}\tilde{H}_{|_{z_e}} = 0$.

For the differential of potential energy we have
\[ \bsym{d}U_{|z_e} = \partial_r U\bsym{d}x_1 + \partial_{c'''} U \bsym{d}\nu_3
\leqno(29)\]

For the differential of kinetic energy we have
\[ \bsym{d}T_{|z_e} = \frac{p_0}{M} \bsym{d}p_2 + \alpha n_0\bsym{d}n_3;
\leqno(30)\]

Collecting the differentials of efficiency function, we get
\[ \bsym{d} \tilde{H}_{|z_e} = (\partial_r U_{|z_e} - \omega p_0)\bsym{d}x^1
  + \left(\frac{p_0}{M} - \omega r_0\right)\bsym{d}p_2
\leqno(31)\]
\[ + (\partial_{c^{'''}} U_{|z_e} - \lambda_1 + \lambda_2 n_0)\bsym{d}\nu^3
   + (\alpha n_0 - \omega  - \lambda_2)\bsym{d}n_3
\]

Equating $\bsym{d} \tilde{H}_{|z_e}=0$, we derive
the  following expression for Lagrange multipliers
\[ \begin{cases}
      p_0/M = \omega r_0;\\
      \omega p_0  = \partial_r U_{|z_e} = \frac{3 K r_0}{R^2};\\
      \lambda_2 = \alpha n_0 - \omega; \\
      \lambda_1 = \partial_{c'''} U_{|z_e} + \lambda_2 n_0
                = K + n_0(\alpha n_0 - \omega),
  \end{cases}
\leqno(32) \]
where
\[ K = \partial_{c'''} U_{|z_e} = \frac{\lambda_0 h}{2\pi R^3}
\leqno(33) \]

the first equation in (32) is an ordinary relationship between
linear and angular velocity during circular orbital motion.

second equation in (32) represents the equality of centrifugal (on the left)
and centripetal (on the right) forces.

From this two expressions we get the relationship for angular velocity, namely:
\[ M\omega^2 = \frac1{r_0} \partial_r U_{|z_e} = \frac{3 K}{R^2}
\leqno(34) \]

\newpage
\subsection{Allowable variations }
\label{Stability2}
\bigskip

For the application of the sufficient condition of stability in the theorem 4.8. in [12]
it is necessary to extract a linear subspace of allowable variations.

Let’s consider the variations of the dynamic variables annihilating
the differentials in formula (27).

From the second line in (27) it follows, that $\delta\nu^3=0$,
then it ensues from the third line, that $\delta n^3=0$.

Thus, we obtain
\[ \begin{cases}
   \delta\nu^3 = 0; \\
   \delta n_3 = 0; \\
   \delta p_2 = -\frac{p_0}{r_0}\delta x_1;
\end{cases}
\leqno(35) \]
Hence it ensues that the variations in the form
\[\delta x^1,\delta x^2, \delta x^3;\quad
  \delta p_1, \delta p_3;\quad
  \delta\nu^1,\delta\nu^2;
  \quad \delta n_1,\delta n_2
\leqno(36)\]
can be considered as independent variations,
furthermore, we must exclude from this subspace
the direction which is  tangent to the orbit

It ensues from formula (18),
that this direction (in $z_e$ point) is determined as
\[ \begin{cases}
   \delta\vec{x} =  r_0\vec{e}_2; \\
   \delta\vec{p} = -p_0\vec{e}_1; \\
   \delta\vec{\nu} =  0; \\
   \delta\vec{n} = 0.
\end{cases}
\leqno(37)\]

In order to eliminate the variation (37),
we impose another additional condition on variations,
and then we get the constraints
\[ \begin{cases}
   \delta\nu^3 = 0; \\
   \delta n_3 = 0; \\
   \delta p_1 = \frac{p_0}{r_0}\delta x_2;\\
   \delta p_2 = -\frac{p_0}{r_0}\delta x_1;
\end{cases}
\leqno(38) \]
and an independent set of variations will be
\[\delta x^1,\delta x^2, \delta x^3;\quad
  \delta p_3;\quad
  \delta\nu^1,\delta\nu^2;\quad
  \delta n_1,\delta n_2;
\leqno(39)\]

\newpage
\subsection{Basic quadratic form }
\label{Quadratic}
\bigskip

Sufficient condition for a minimum consists in positive definiteness of quadratic form
of type $\bsym{d}^2\tilde{H}_{|_{z_e}} (\delta z, \delta z^{'})$,
where variation vectors $\delta z$,~$\delta z^{'}$
must be expressed through independent variations (39)
taking into account the constraints (38).
Quadratic form defined in independent variations we denote by $Q$.

Calculations of the efficiency function hessian (adjoined Hamiltonian)
and basic quadratic form in independent variations were performed in Maple.

For better structuring of the expressions indefinite
Lagrange multipliers are hidden at the first stage.

After insignificant transposition of columns (and corresponding lines with the same number)
the matrix of basic quadratic form acquires a form
\[ \begin{bmatrix}  Q_{11} & 0 & 0 & 0 & 0 & 0 & 0 & 0 \\
                    0 & Q_{22} & 0 & 0 & 0 & 0 & 0 & 0 \\
                    0 & 0 & Q_{44} & 0 & 0 & 0 & 0 & 0 \\
                    0 & 0 &      0 & Q_{33} & Q_{35} &   0 & 0 & 0 \\
                    0 & 0 &      0 & Q_{35} & Q_{55} & Q_{57} & 0 & 0 \\
                    0 & 0 &      0 & 0      & Q_{57} & Q_{77} & 0 & 0 \\
                    0 & 0 & 0 & 0 & 0 & 0 &   Q_{66} & Q_{68} \\
                    0 & 0 & 0 & 0 & 0 & 0 &   Q_{68} & Q_{88}
    \end{bmatrix}
\leqno(40)\]

Lets write out non zero elements from matrix of quadratic form (per line)
\[ Q_{11} = 3\left(\frac{h^2-4 r_0^2}{R^2}\frac{K}{R^2}
          + M \omega^2\right)
\leqno(41)\]
\[ Q_{22} = 3\left(\frac{K}{R^2} + M \omega^2\right)
\leqno(42)\]
\[ Q_{44} = \frac{1}{M}
\leqno(42)\]
\[ Q_{33} =  \frac{3 r_0^2 - 2 h^2}{R^2}\frac{3 K}{R^2},\qquad
   Q_{35} = -\frac{3 K r_0}{R^2}
\leqno(43)\]
\[ Q_{55} = \lambda_1,\quad
   Q_{57} = \lambda_2
\leqno(44)\]
\[ Q_{77} = \alpha
\leqno(45)\]
\[ Q_{66} = \lambda_1 = Q_{55},\quad
   Q_{68} =\lambda_2 = Q_{57}
\leqno(46)\]
\[ Q_{88} = \alpha = Q_{77}
\leqno(47)\]

Substituting $M\omega^2$ for expression (34) in $Q_{11},Q_{22}$, we obtain
\[ Q_{11} = \frac{3 K}{R^2}\frac{4 h^2 - r_0^2}{R^2}
\leqno(41a)\]
\[ Q_{22} = \frac{12 K}{R^2}
\leqno(42a)\]

\newpage
\subsection{Conditions of positive definiteness of the basic quadratic form }
\label{Definite}
\bigskip

For matrix $Q$ to be positive definite it is foremost necessary
that all diagonal elements of the matrix are positive.
$Q_{22},Q_{44},Q_{77},Q_{88}$ are scienter positive.
Remaining conditions are as follows
\[ \begin{cases}
      0 < Q_{11} = \frac{3 K}{R^2}\frac{4 h^2 - r_0^2}{R^2};\\
      0 < Q_{33} = \frac{3 K}{R^2}\frac{3 r_0^2 - 2 h^2}{R^2};\\
      0 < Q_{55} = Q_{66} = \lambda_1 = K + n_0(\alpha n_0 - \omega);\\
  \end{cases}
\leqno(48) \]

The first two conditions result in purely geometrical limitations
\[ \left(Q_{11} > 0\right) \& \left(Q_{33} > 0\right)\longrightarrow
  \left(\sqrt{\frac23}< \frac{r_0}{h}\right) \& \left(\frac{r_0}{h} < 2\right)
\leqno(49) \]

These conditions of positive definiteness of the matrix $Q$
have to be supplemented now by the conditions of positive definiteness
of two submatrices of $3\times 3$ and $2\times 2$, namely
\[ \begin{bmatrix}
                    Q_{33} & Q_{35} &   0 \\
                    Q_{35} & Q_{55} & Q_{57} \\
                    0      & Q_{57} & Q_{77}
   \end{bmatrix}
\leqno(50)\]
and
\[ \begin{bmatrix}
                    Q_{66} & Q_{68} \\
                    Q_{68} & Q_{88}
    \end{bmatrix}
\leqno(51)\]

Taking into account the above mentioned considerations it is sufficient for the matrix (51)
to check the condition of positiveness of its determinant of $Q_{66} Q_{88} - Q_{68}^2$.
Thus, additionally to the conditions (49) the following condition is added
\[ 0 < Q_{66} Q_{88} - Q_{68}^2 = \alpha K + \omega(\alpha n_0 - \omega)
\leqno(52) \]

Now we investigate the conditions of positive definiteness of the matrix (50).
The first condition $Q_{33}>0$ we have considered already.

Thus the additional conditions of positive definiteness of matrix (50)
are reduced to positiveness of two determinants
\[ 0 < Q_{33} Q_{5,5} - Q_{35}^2
\leqno(53) \]
and
\[ 0 < Q_{33} Q_{55} Q_{77} - Q_{33} Q_{57}^2 - Q_{77} Q_{35}^2
\leqno(54) \]

Condition (54) can be also written in form
\[ Q_{77} (Q_{33} Q_{55} - Q_{35}^2) > Q_{33} Q_{57}^2
\leqno(54a) \]
and, since $Q_{77} > 0$ then (54) transforms into (54b)
which replaces condition (53) as it accounts for it
\[ Q_{33} Q_{55} - Q_{35}^2 > \frac{Q_{33}}{Q_{77}}Q_{57}^2
\leqno(54b) \]

So, condition (53) is superfluous and it is necessary
to study only condition (54). Remind that
\[ \begin{cases}
      Q_{33} = \frac{3 K}{R^2}\frac{3 r_0^2 - 2 h^2}{R^2};\\
      Q_{35} = -3\frac{K r_0}{R^2};\\
      Q_{55} = K + n_0(\alpha n_0 - \omega) = Q_{66}; \\
      Q_{57} = \alpha n_0 - \omega = Q_{68}; \\
      Q_{77} = \alpha = Q_{88};
  \end{cases}
\leqno(55) \]

Now write down (54) in form
\[ Q_{55} Q_{77} - Q_{57}^2 > \frac{Q_{77} Q_{35}^2}{Q_{33}}
\leqno(54c) \]
as supposed $Q_{33}>0$.

Taking into account formulas (55), condition (54c) is equivalent to
\[ Q_{66} Q_{88} - Q_{68}^2 > \frac{Q_{77} Q_{35}^2}{Q_{33}} > 0
\leqno(54d) \]

Thus, condition (52) is a consequence of condition (54) and $Q_{33}>0$ from (48).
It means that condition (52) can be omitted.

We have the following reduced number of conditions for matrix $Q$ positive definiteness:
\[ \begin{cases}
      0 < Q_{11} = \frac{3 K}{R^2}\frac{4 h^2 - r_0^2}{R^2};\\
      0 < Q_{33} = \frac{3 K}{R^2}\frac{3 r_0^2 - 2 h^2}{R^2};\\
      0 < Q_{55} = Q_{66} = \lambda_1 = K + n_0(\alpha n_0 - \omega);\\
      0 < Q_{33} Q_{55} Q_{77} - Q_{33} Q_{57}^2 - Q_{77} Q_{35}^2
  \end{cases}
\leqno(56) \]

We investigate condition (54) in form
\[ 0 < Q_{33} (Q_{55} Q_{77} - Q_{57}^2) - Q_{77} Q_{35}^2
\]
\[
 = Q_{33} (Q_{66} Q_{88} - Q_{68}^2) - Q_{77} Q_{35}^2
\]
\[ = \frac{3 K}{R^4}[(3 r_0^2 - 2 h^2)\omega(\alpha n_0 - \omega) - 2\alpha K h^2]
\]
That is
\[ \frac{\omega}{\alpha}(\alpha n_0 - \omega) > K \frac{2 h^2}{3 r_0^2 - 2 h^2}
\]

So, condition (54) is equivalent
\[ \frac{\omega}{\alpha}(\alpha n_0 - \omega) > \frac{K}{\frac32 (\frac{r_0}{h})^2 - 1}
\leqno(57)\]

In particular, as $Q_{33}>0$, we have \( \alpha n_0 - \omega > 0\), and it means
that conditions $Q_{55}=Q_{66}=\lambda_1>0$ are fulfilled a priory and can be omitted.

Therefore, conditions
\[ \begin{cases}
      \sqrt{\frac23}< \frac{r_0}{h} < 2;\\
      \frac{\omega}{\alpha}(\alpha n_0 - \omega) > \frac{K}{\frac32 (\frac{r_0}{h})^2 - 1}
  \end{cases}
\leqno(58) \]
define positive definiteness of form $Q$.

\newpage
\subsection{Physical meaning of positive definiteness conditions }
\label{Definite2}
\bigskip

In the previous section we have established the conditions of the system’s parameters
which provide positive definiteness of basic quadratic form, namely:
\[ \begin{cases}
      \sqrt{\frac23}< \frac{r_0}{h} < 2;\\
      \frac{\omega}{\alpha}(\alpha n_0 - \omega) > \frac{K}{\frac32 (\frac{r_0}{h})^2 - 1}
  \end{cases}
\leqno(59) \]

The first condition in (59) is purely geometrical
and determines a possible range for a radius of the orbit
representing a relative equilibrium (21).

The second condition is dynamic and determines
lower boundary for the intrinsic moment of momentum of the body.
In particular, it means that a body must be sufficiently rapidly revolved.
Therefore it is worth to solve this inequality relative to $n_0$.

Using relationships (34), we obtain
\[  n_0 > \frac{\omega}{\alpha }
     + \frac{1}{3}\frac{1+ (\frac{h}{r_0})^2}
     {\frac32 (\frac{r_0}{h})^2 - 1}(\omega M r_0^2)
\leqno(60)\]

Value of $\frac{\omega}{\alpha }$ corresponds to the intrinsic
moment which a body would have if it were to revolve with angular velocity
$\omega$ athwart to the own axis of symmetry.

Value of $\omega (M r_0^2)$ is simply the orbital moment of momentum $(L_z)_{|z_e}$.

For condition (60) one can give such physical meaning:
{\it intrinsic moment of body rotation must be of the same or higher order of its orbital moment}.

Indeed, multiplier before $\omega (M r_0^2)$
is a geometrical factor which is $\frac{r_0}{h}=1.0$
equal to $\sim 1.33$, and at $\frac{r_0}{h}=1.5$ equal to $\sim 0.2$.

In these estimations the first term in the right part of expression (60) can be neglected.

\newpage
\section{Numeral simulation }
\label{numeric}
\bigskip

From a mathematical point of view the stability conditions obtained here
allow a wide range of parameters values of the problem to exist,
however not all of them can be physically realized.
From a physical perspective the values of parameters are limited by
the properties of present materials.

In addition it seems difficult to realize in practice high
rate of rotations, especially as far as it concerns intrinsic angular velocity
of movable magnetic body (dipole).

Therefore it appears necessary to specify such values
of parameters which can be realized in an experiment.

For the magnets, made from $Nd-Fe-B$, we have the following
characteristics: $\rho = 7.4\cdot 10^3(kg/m^3)$ – density
and $B_r = 0.25 (T)$ -- remaining induction.
Then it is easy to obtain magnetic ‘‘charge’’ of the poles $\kappa = 17.6 (A \cdot m)$.
Distance between the poles is $L = 2h = 0.1(m)$.

A movable magnet we choose in a form of cylinder (disk)
with the diameter of $d = 0.014(m)$ and height $l = 0.006(m)$.
Then disk magnetic moment $\mu = 0.18 (A \cdot m^2)$.

As a result for the orbit with the radius $r_0 = 1.5 h = 0.075 (m)$ we obtain the angular velocity
of the orbital motion $\omega = 1.54 (rad/sec)$, with minimum angular velocity
of disk intrinsic rotation in this case is $\Omega = 72.8(rad/sec)$.
Such values of angular velocity appear fully reasonable.

Using the indicated values of Orbitron parameters
the numeral modeling of orbital motion was conducted
under the deviations of initial values of dynamic variables
from the values, which correspond to relative equilibrium within $1\%$ error.
1000 castings which showed the stability of orbital motion were accomplished by Monte Carlo method
(i.e. by random selection of the initial values in the vinicity).
We will elucidate this in more detail in the next parts of this work.

\newpage
\section{Summary}
\label{summary}
\bigskip

The main aim of this work was to give constructive proof
of stable orbital motions existence in the systems of bodies,
which interact only by magnetic forces.

For this purpose it is enough to analytically prove the existence of stability
for one orbit in comparatively simple system described by equations which do not contradict
the laws of electrodynamics and classical mechanics.

We named such a system Orbitron,
found its parameters which can be physically realized
and conducted the Monte Carlo numeral modeling.
\bigskip

{\it To be continued} $\dots$

\bigskip
\section{References}
\label{literature}
\begin{enumerate}
\linespread{2.5}
\item \textit{Zub\,S.\,S.} in Proceedings of the Int. Conference on Magnetically Levitated Systems and Linear Drivers (MAGLEV'2002), Lausanne, Switzerland, 2002, eConf CPP02105 (2002).
\item \textit{Zub\,S.\,S.} Influence of superconductive elements topology on the stability of the free body equilibrium, (Ukrainian) / S.S.~Zub // Synopsis of Ph.D. Dissertation, Institute of Cybernetics, National Academy of Sciences of Ukraine, Kiev, 24 p. 2005.
\item \textit{Ginzburg\,V.\,L.} Mezotrons Theory and Nuclear Forces / V.L.~Ginzburg // -- Phys.-Uspekhi. --1947. -- Vol.~31., issues 2. -- P. 174 -- 209.
\item \textit{Schwinger\,J.} A Magnetic Model of Matter, Science 165 (No. 3895), 757 (1969).
\item \textit{Harrigan\,R.\,M.} Levitation device, U.S. Patent 382245, May 3, 1983.
\item \textit{Kozoriz\,V.\,V.} About a problem of two magnets / V.V.~Kozorez // Bull. of the Ac. of Sc. of USSR, Mech. of a Rigid Body. -- 1974. -- N3. -- P. 29 -- 34.
\item \textit{Kozoriz\,V.\,V.} Dynamic Systems of Free Magnetically Interacting Bodies, (Russian) / V.V.~Kozoriz // Naukova Dumka, -- Kyiv, 1981. -- 139 p.
\item \textit{Zub\,S.} Research into Orbital Motion Stability in System of Two Magnetically Interacting Bodies, [math-ph/1701], arXiv:1101.3237
\item \textit{Zub\,S.\,S.} Research into orbital motion stability in system of two magnetically interacting bodies / S.S.~Zub // Visnyk Taras Shevchenko KNU. --- Physics and Mathematics. -- 2011. Vol. 2. -- P. 176 -- 184.
\item \textit{Marsden\,J.\,E.} Introduction to Mechanics and Symmetry / Jerrold~E.~Marsden, Tudor~S.~Ratiu // Cambridge University Press, -- London, 1998. -- 549 p.
\item \textit{Marsden\,J.\,E.} Lectures on Mechanics. -- London : Cambridge University Press, 1992. -- 254 p.
\item \textit{Ortega\,J-P., Ratiu\,T.\,S.} Non-linear stability of singular relative periodic orbits in Hamiltonian systems with symmetry // J. Geom. Phys. -- 1999. -- 32. -- P. 160 --188.
\item \textit{Marsden\,J.\,E.} Hamiltonian reduction by stages / Marsden J.E., Misiolek G., Ortega J.P. et al. // Springer, -- Berlin,  2007. -- 519 p.
\item \textit{Zub\,S.\,S.} Hamiltonian formalism for magnetic interaction of free bodies / S.S.~Zub, S.I.Lyashko // J. Num. Appl. Math. -- 2012. issues 2(102). --P. 49 -- 62.
\item \textit{Simon\,M.\,D.} Spin stabilized magnetic levitation / M.D.~Simon, L.O.~Heflinger, and S.L.~Ridgway // Am. J. Phys., 65, 286–292 (1997).
\item \textit{Zub\,S.} Mathematical model of magnetically interacting rigid bodies // PoS(ACAT08)116. -- 2009. -- 5 p.
\item \textit{Tamm\,I.\,E.} Fundamentals of the theory of electricity / I.E.~Tamm // Mir, (1979).
\item \textit{Landau\,L.\,D.} Electrodynamics of continous media / L.D.~Landau, E.M.~Lifshitz // Pergamon, (1960).
\item \textit{Smythe\,W.\,R.}, Static and Dynamic Electricity / W.R.~Smythe // McGraw-Hill, New York (1939).
\end{enumerate}

\end{document}